# Light and Shadows over Petra: astronomy and landscape in Nabataean lands

**Juan Antonio Belmonte**
Instituto de Astrofísica de Canarias, La Laguna, Tenerife, Spain.
Departamento de Astrofísica, Universidad de La Laguna, Tenerife, Spain.
jba@iac.es

**A. César González-García**
Instituto de Astrofísica de Canarias, La Laguna, Tenerife, Spain [former address].
Instituto de Ciencias del Patrimonio, Incipit, CSIC, Santiago de Compostela, Spain.
cesar.gonzalez-garcia@incipit.csic.es

**Andrea Polcaro**
Università degli Studi di Perugia, Perugia, Italy.
andrea.polcaro@gmail.com

**Abstract.**

The Nabateans built several monuments in Petra and elsewhere displaying a decoration with a preference for astronomical motifs, possibly as a reflection of their religion. However, due to the lack of direct written accounts and the scarcity of inscriptions we do not have a clear knowledge on the precise nature of such believes and how these reflected on the calendar or the religious time-keeping system of this ancient society. A statistical analysis of the orientation of their sacred monuments demonstrates that astronomical orientations were often part of an elaborated plan and possibly a trace of the astral nature of Nabataean religion. Petra and other monuments in the ancient Nabataean kingdom have proven to be marvellous laboratories of the interaction between landscape features and astronomical events showing impressive hierophanies on particular monuments related to cultic times and worships. Among other findings, the famous Ad Deir has shown a fascinating ensemble of light and shadow effects, perhaps connected with the bulk of Nabataean mythology, while from the impressive Urn tomb, a series of suggestive solstitial and equinoctial alignments emanate which might have lately helped its selection as the cathedral of the city. This paper demonstrates that the sky was a substantial element on Nabataean religion and reveals new evidence for cultic worship centred on the celestial sphere.

*1. Introduction*

From the victory of the legendary king Obodas I upon Seleucid armies in the early first century BC to the annexation of their kingdom to the Roman Empire by Emperor Trajan at the beginning of the 2$^{nd}$ Century AD –and even later–, the Nabataeans, a people of presumable Arab lineage, developed a singular and sophisticated culture in the harsh lands of Arabia Petraea (to the southeast of Palestine and Syria in antiquity) at the frontiers of the Hellenistic World [Markoe 2003, Bowersock 2003]. For centuries, they carved out hundreds of tombs in the illustrated sandstone and built palaces for their kings and temples for their divinities, creating one of the most fascinating places on Earth, the legendary city of Petra, Nabataean Reqem. This was the capital of their kingdom for generations and represented the highlight of their civilization although the Nabataean genius was also present in many other sacred buildings scattered across their lands (see Figure 1). Among these, the nearly contemporaneous temples at Khirbet et Tannur and Khirbet ed Dharih do show a collection of elements of undoubted astral symbolism which might be traced in the nature of Nabataean religion [Gawlikowski 1990]. In particular, Tholbecq [1997] has convincingly suggested that certain busts uncovered at the excavations of Dharih actually represent the seven planets.

There are indications that the Nabataeans had a naturalistic religion, a strange mixture of elements from pre-Islamic Arabs and Hellenistic, Egyptian and other Middle Eastern influence [Healy 2000]. Stone blocks (*baetyles*) often represented the divinities although human or quasi-human forms were developed lately. The principal male divinity was the god Dushara or Dushares, very probably an astral god with a hypothetical lunar or solar character. His name means "He of Shara", Shara being the mountain range to the east of Petra where the ancient site of Gaia (the presumable previous capital city) is located. There are evidences of a certain syncretism with the god Al−Kutba, meaning the "writer" and, consequently, with the Babilonian god Nabu [Gawlikowski 1990]. In this case, the planet Mercury would have been one of his celestial manifestations [Pettinato 1998]. Dushara has been identified either with Greek Zeus, Dionysos or Ares, although the later is a less common association. Archaeological evidence suggests that he was a deity related to the cult of the deceased.

However, there has been much discussion regarding the head female divinity of the Nabataeans pantheon. In Bosra, the northern Nabataean capital during the crepuscular reign of king Rabel II (71-106 AD), the main goddess was Allat, meaning simply "the Goddess". With a hypothetical solar character (As Sams, the Sun, was a female divinity in pre-Islamic Arabia), she has been identified with Athena or Atargatis [Healy 2000]. Her name is also present in Wadi Ramm (ancient Iram). However, in Petra, this name is never found whilst a lot of inscriptions in Nabataean script have been found mentioning the goddess Al Uzza. Her name means "the Most Powerful" and she was the personification of the Evening Star, the planet Venus, identified with the Greek Aphrodite and the Cananaean Astarte and also with the Egyptian goddess Isis. In Nabataean inscriptions found in Petra and Iram, she is mentioned in close relationship with Al Kutba and, of course, Dushara but we still do not know the exact ties between them. Indeed, in the area of Petra, Dushara and Uzza undoubtedly were at the head of a pantheon with many levels of comprehension [Zayadine 2003].

*2. Statistical analysis of orientations*

In December 2011, deliberately in coincidence with the winter solstice, our team, comprising of two archaeoastronomers, specialists in ancient Mediterranean cultures and statistical analyses of series of data [see, e.g. González-García and Belmonte 2011], and an archaeologist working in the Levant, notably in Jordan and Syria [see, e.g. Polcaro 2012], moved to Jordan with the aim of performing a primeval archaeoastronomical analysis of Nabataean monuments in the region. Our goal was to analyse a statistically significant sample of temples and other sacred buildings, which could permit archaeological confirmation of suspected astronomical activities by the Nabataeans relating to religious practice [Belmonte 1999]. Figure 1 illustrates the sites from which data have been collected (including the city of Petra itself), while Table 1 shows the raw data for the different sites and monuments. The sample includes the datum of the commemorative temple of king Obodas at Avdat obtained during a visit to the Negev region in 2008.

The data sample includes 92% of the temples known, including those in Petra and in other Nabataean settlements of the kingdom such as el Qsar, Dhat Ras, Tannur, Dharih or Wadi Ramm (fig. 2). In Petra, data includes temples plus the majority (~80%) of the accessible highplaces (open-air altars carved on the rock in the top of cliffs and conspicuous mountains), including the best known at Djebel Madbah (fig. 3, panel a), and a few of the most representative monuments excavated and sculpted in the sandstone walls. Although the number of rock-cut chambers present in the last census of the city is quite high [Lehme 2003] not all of them had a marked religious character. Our intention was selecting those architecturally significant for which a religious character behind its mortuary use has been definitely proven such as Ad Deir or Monastery, the Urn tomb or the most controversial of them, Al Khazna or Treasury [Stewart 2003]. In total, our data includes fifty temples and other cultic structures from all over the ancient Nabataean kingdom that we estimate to be a statistically significant sample of all known religious structures up to date.

Data were collected using high precision compasses and clinometers and corrected for magnetic declinations. Magnetic alterations are not expected in the Nabataean territory, where most of the terrain is limestone or sandstone. The measurements included in Table 1 have an average error of ¼° in azimuth and ½° in horizon altitude which translates into an error ~¾° in declination. Figure 4 illustrates the results of our work. Panel a shows the orientation diagram of the sample in azimuth where it can be observed that the axes of most structures are concentrated either in the solar arch or in a general orientation towards the meridian. Figure 4 (panel b) shows the astronomical declination histogram, a magnitude independent of geographic coordinates and local topography. The declination histogram was calculated using a density distribution with an Epanechnikov kernel with a pass band of 1½°. This histogram is similar to the one discovered for neighbouring cultures with a strong astral component in their religion such as ancient Egypt [Belmonte, Shaltout and Fekri 2009] and shows a series of significant peaks. Significance is estimated by the following procedure: the mean is first computed and subtracted from the data. Then, the data are normalized with the standard deviation of the measurements. Any peak rising above the 3σ level could be considered as having a degree of confidence higher than 99% within this particular significance test.

Some of the peaks of the histogram, of a probable astral –presumably solar– character, might be interpreted at the light of Nabataean beliefs, considering, among other sources, that Strabo[1] reported that the Nabataeans worshipped the Sun on the roof of their houses. Peak I, centred at −0¼° could indeed be catalogued as equinoctial. Scholars have suspected that the temple of Tannur may be associated with the equinoxes [Villeneuve and Al-Muheisen 2003, Mckenzie 2003], a moment for presumable pilgrimages to the top of the mountain where the temple is located (see fig. 2), a fact hardly surprising considering the abundance of astral symbolism in the sculpture rescued at Tannur and the neighbouring temple of Dharih. This may suggest that the equinoxes were important marks in the Nabataean sacred time and a possible way to control time within the framework of a lunisolar calendar. The results of this study confirm this suspicion since there is a significant general trend in the data to the time frame of the equinoxes (declination 0º). The Urn tomb and the Obelisks at Djebel Madbah are also relevant in this context, as we will see below. Peak II at 24¼° is certainly solstitial while peak III, centred at −25¼° could be related with any of the celestial bodies moving close to the Ecliptic and that were relevant in Nabataean religion: the winter solstice sun (according to Epiphanius's *Panarion*), Venus or Mercury, as stated above. For the additional peak above the 3σ level within the solar range in fig. 4, at a declination of ~8°, there is however no direct or simple astronomical explanation.

To understand peaks IV and V we must take into account that the Nabataeans were a people of presumable Arab lineage. Peak IV, centred at 60¼°, is certainly the accumulation peak to northern directions related to the average latitude of the Nabataean Kingdom: 30ºN. This could be connected with the large number of monuments which were northerly orientated, including the main temples at the colonnade avenue in Petra. One of those temples was the singular Qsar el Bint, arguably the main sanctuary of Dushara, situated at the convergence of the main caravan roads leading to the city centre which was possibly founded by king Obodas III (28-9 BC), and completed by Aretas IV. It was severely damaged by an earthquake (perhaps the one nearly destroying the city in 363 AD) but recent excavations have recovered part of the starry decoration in stucco [Larche and Zayadine 2003]. An additional, although less significant peak (V) is located at a declination of −52¾°, very close for the epoch to the declination of the bright star Canopus, which could however be interesting in a most general context.

According to Arabic sources of the early Muslim era, the Ka'aba in Mekka had a main axis orientated to Suhail, the Arabic for Canopus, and the stars of the Handle of the Plough (Alkaid had a declination of 60º c. 1 AD) and a minor axis oriented according to the solstitial line [Hawkins and King 1982]. The black stone was embedded in the SE corner of the monument facing the equinoxes. It is certainly curious that some Nabataean monuments do reproduce the same pattern of alignments as those classically reported for a hypothetical pre-Islamic Arabic temple such as Ka'aba. The last surviving inscription in the Nabataean language dates from 356 A.D. a quarter of a millennium earlier than the arrival of Islam but the Nabataean divinities were certainly worshiped in the region –the sanctuary of Al Uzza at Wadi Hurad was destroyed by Khalid Ibn al Walid at the commandment of Mohammed immediately after the capture of Mekka–. It is worth noticing that "by Allat, Al Uzza and Manat, the third of the triad" Kuraish performed "tawat" in Ka'aba [Zayadine 2003], although the impossibility of handling archaeological excavation in the Holly Mosque disables any further conclusions. However, most of the peaks in the histogram can be interpreted at the light of Nabatean beliefs reinforcing the astral character of the religion and showing that the equinox and perhaps the solstices were important for their time-keeping.

Peaks IV and V in fig. 4 (panel b) correspond to the accumulation peaks related to northern and southern azimuths. These correspond to areas where a large number of declinations are available within a certain azimuth interval and consequently some of the peaks towards these directions could have a spurious nature. We have been puzzled by this possibility since our research group started statistical analysis of temple orientation in the Mediterranean region but was difficult to prove with samples of a high variability in latitude and horizon angular heights. In order to test this phenomenology in a compact geographical area, we have performed a preliminary test of our Nabataean data by comparing the distribution of declinations of our sample with that arising from a homogeneous set of orientations with the same population. Figure 5 shows the result of this comparison after subtracting such distribution and normalizing by the standard deviation of the homogeneous sample. We can see that the vast majority of the relevant peaks are still present, including peak IV towards northern declinations which is certainly significant. However, peak V is absent and seems compatible with a homogeneous distribution. Consequently, it must be considered with more caution. This test is still a preliminary trial and must indeed be refined. It does not include the effect of varying horizon altitudes and at present state may not work well for larger geographical areas such as ancient Egypt. However, it behaves reasonably well for the Nabataean realm and the results seem robust.

*3. Light and shadow effects*

Our campaign in Nabataea intended to observe the effect of the winter solstice phenomena at some of the most impressive monuments of Petra. Belmonte [1999] suggested a phenomenology related to the solstices for some of the most singular monuments in Petra, such as the Monastery or the Treasury. Direct observation would enable us to directly witness light and shadow effects that may have been of significance to the Nabataeans.

The most impressive light and shadow effect at winter solstice occurs at Ad Deir (the Monastery). It is unclear if this is the temple of one of the most important Nabataean divinities, Dushara or Uzza, a *heroon* for one of their deified kings such as Obodas I or the unfinished burial place or cenotaph of one of their last kings such as Rabel II. Its use as a church in the Byzantine era and its internal distribution suggests that, originally, this was a sort of monumental cella or biclinium with a cultic podium (a môtab) on its back [Wenning 2003]. Indeed, Ad Deir would have been a prominent festival venue, with an elaborated staged ascent from the centre of the city, a vast court in front of it and a series of related monuments such as a stonecircle, an altar and the temple–like building known as structure 468 situated in front of it. The orientation of the structure, shown in Table 1, and especially in Figure 6, strongly suggests a winter solstice relationship.

On the one hand, the left image of fig. 6 shows the light and shadow effect produced at the interior of the monument at the moment of winter solstice sunset. The light of the setting sun entering through the gate of the monument perfectly illuminates the sacred area of the deep interior of the building where the môtab for the installation of the sacred baetyls is located. The effect is spectacular and would have been observable only a week or so before and after the winter solstice. On the other hand, winter solstice sunset, as observed from the môtab itself, is produced in a most peculiar way on a rock with the aspect of the head of a lion –the sacred animal of Al Uzza– as shown in the right panel of fig. 6. At present time the sun sets at least twice, first in the axis of the monument and then re-appears in the northernmost corner of the rock before its final disappearing. The phenomenon would have been still more impressive two thousand years ago when the northern limb of the disk of the sun had a declination close to $-23½°$. We believe that this extraordinary ensemble of solar hierophanies, perhaps in combination with the visibility after sunset of other celestial bodies such as the Evening Star, clearly reinforces the idea of the Monastery as one of the most important sacred enclosures of the Nabataean realm. Ad Deir would have been the ideal place to celebrate, in dates close to the winter solstice, the birth of Dushara from his own mother-cum-consort Al Uzza, the goddess of fertility.

As mentioned above [Villeneuve and Al-Muheisen 2003], the knowledge of the equinoxes was of particular importance to the Nabataeans, and could have been a key element for the control of a lunisolar calendar. Interestingly, our new data confirm the equinoctial alignment of the impressive Zibb Attuf, the "Pillars of Merciful" (see fig. 3, panel b) popularly known as the Obelisks. These carved out behemoths could have been used to control time by the use of shadow casts at sunrise. However, in the mid-1990s [Belmonte 1999], the most inspiring equinoctial relationship had been suggested for the Urn Tomb –the most impressive and better preserved of the so-called royal tombs at the western cliffs of Djebel Khubtha (see fig. 3, panel c)– and the impressive mountain of Umm al Biyara (see fig. 1). This sacred mountain was very important for the Nabataeans not only due to its unassailability but also because it was the main source of water for the city (the Siyagh Spring, the only large permanent water supply in Petra, was located at its base). The royal tombs seem to have been built in that sector of Djebel al Khubtha where sunset at the equinoxes was visible over the top of Umm al Biyara. For example, the well-preserved gate of the Urn Tomb was centred at equinox sunset over the central part of that particular mountain. Our new data (see Table 1) plainly confirm this earlier result but to a much larger degree of sophistication.

This is demonstrated in Table 1 and Figure 7. The Urn tomb has a quite elaborate design, with a large court in front of the structure and a big hall excavated on the sandstone of the cliff, suggesting that it was used not only as a tomb −perhaps of king Malichus II− but also as a place for other religious activities or festivals, possibly related to the cult of the dead, in apparent connection to the Autumnal equinox. On December 21$^{st}$ 2011, sunset at the winter solstice was observed from the court in front of the Urn Tomb. During the sunset, the sun passed behind a conspicuous landmark in the distant western horizon. Most important, the last rays of the sun illuminated the northeast corner of the inner hall after crossing the main gate of the tomb. The phenomenon, in combination with the confirmed equinoctial alignment (see fig. 7), proved quite astonishing.

Sunset at the equinox took place between two distinct features on the summit of Umm al Biyara.[2] Surprisingly, our measurements (see Table 1) also indicate that sunset at the summer solstice occurs in between another couple of this kind of "natural" features further to the north in the distant western horizon and that this new alignment completes the symmetry of the main hall of the tomb (see fig. 7). This impressive set of three alignments within the plan of the tomb in combination with significant features in the distant horizon can hardly be ascribed to chance. We argue here in favour of a deliberate attempt to convert the hall of the Urn tomb, whatever its actual purpose −certainly religious−, in a kind of time-keeping device that would have been very useful in controlling time and the calendar, be it sacred or profane. We may conclude that this is the result of an original Nabataean design, considering the findings in other buildings commented throughout the paper.

Interestingly, Bishop Jason converted the Urn tomb into the cathedral church of Petra on June 24, 446 AD [Fiema 2003]. We may thus suggest that this formidable enclosure, indeed a place with a sacred character, was selected as the new cathedral of the city because it included such notable grouping of alignments, so useful for the Christian worship. The three alignments would have offered markers, of an excellent and precise nature, for the determination of Christmas, on December 24, Easter (through the observation of the spring equinox), and Saint John, on June 24, precisely the date of consecration of the new cathedral.

Finally, Al Khazna, or Treasury, is possibly the nicest monument ever erected by the Nabataean kings. Located at a breath-taking position at the exit of As Siq, the narrow canyon that approaches the city from the east, it is the first large building that a visitor to Petra faces when entering the city through Siq (see fig. 3, Panel d). The discovery in recent excavations of what seems to be a couple of traditional burial chambers cut in the rock just below the impressive façade of the Treasury[3] has reinforced the idea that this magnificent monument was something more than a simple royal tomb. This fact suggests that, even if it was the tomb of an important king, such a Aretas IV, or the *heroon* of Obodas I −both alternatives are the most popular−, it might have acted as a sort of sanctuary where the genius protector of the city, a manifestation of Tyche represented in the façade of the building, would have been venerated. In the well-known zodiac-stone found at the temple of Tannur, Tyche is represented dominating the scene and associated with a crescent of the moon (see fig. 8). It is worth mentioning that the large cliff enclosing the Siq only permits to be seen a very small section of sky at ~18° of angular height from the *sancta sanctorum* of the monument, thus breaking any solar alignment (see Table 1). However, bearing in mind these considerations, it is possible that lunar events could be compatible with the orientation and layout of the internal structure and the external decoration of the monument. This fact, provided it could be proven in the future, would reinforce the sanctuary nature of the building.

*4. Conclusion*

The statistical analysis of our sample of data, together with the analysis of the light and shadow effects confirmed in several monuments of the city related to the consistent use of the equinoxes, the solstices and perhaps other conspicuous astronomical features, undoubtedly points towards the importance of astral elements in Nabataean religion. These events could have been used to mark times of worship and, most important, to control a calendar, and certainly convert the city of Petra −a place of *awe-inspiring crystallization of natural beauty* and *the unique artistic creation of the Nabataean will*− in *a gift from their gods, shaped by the supernatural and holding a holly meaning* [Jowkowski 2003], that our work has started to unveil.

*Acknowledgments*

We thank Dr. Efrosyni Boutsikas for critical discussions and reading of the manuscript. This work is partially financed under the framework of the projects P310793 "Arqueoastronomía" of the IAC, and AYA2011-26759 "Orientatio ad Sidera III" of the Spanish MINECO.

*Notes*

1. STRABO, *Geographia* XVI, 4, 26.
2. It is not easy to ascertain if these two elements are purely natural, artificial, or natural but re-elaborated. A closer inspection of the mountain summit would be needed.
3. These excavations in the Khazna Courtyard were carried out by the Petra Archaeological Park (PAP) in the years 2003, 2004, 2005, and 2007. However, to our knowledge, no report of these findings has been published yet.

*About the authors*

Juan Antonio Belmonte is a staff astronomer at the Instituto de Astrofísica de Canarias (Tenerife, Spain) where he has lectured history of astronomy and archaeoastronomy and investigates in exoplanets, stellar physics and cultural astronomy. He has published or edited a dozen books and authored nearly 200 publications on those subjects. He has been the Director of the Science and Cosmos Museum of Tenerife from 1995 to 2000 and President of the European Society for Astronomy in Culture (SEAC) from 2005 to 2011. He is advisory editor of the Journal for the History of Astronomy. In the last years he has been performing extensive research on the astronomical traditions of ancient civilizations, concentrating in the ancient Mediterranean cultures. Born in Murcia (Spain) in 1962, he studied physics and got his master-thesis in 1986 at Barcelona University and obtained his PhD on Astrophysics at La Laguna University in 1989.

Antonio César González García (Valladolid, 1973), PhD in Astrophysics (Groningen, The Netherlands) has held postdoctoral fellowships at the IAC (2003-2006 & 2010-2011) and the Theoretical Physics Department – UAM (2006-2010), where he has investigated in evolution of galaxies and archaeoastronomy (working on the possible astronomical orientation of megalithic monuments in central Europe). Since 2010 he is holding a *Ramón y Cajal* fellowship to work on cultural astronomy of Mediterranean cultures. He is now based at the Instituto de Ciencias del Patrimonio (Incipit-CSIC) at Santiago de Compostela (Galicia, Spain). He is vice-President of the European Society for Astronomy in Culture since 2011. His main research lines are centred in three issues: (i) modelling of the possible astronomical orientation of classical cultures, (ii) possible astronomical and landscape relations of Iron Age sanctuaries and (iii) the study of the orientation of ancient Roman cities.

Andrea Polcaro is contract professor of *Archaeology of the Ancient Near East* at the Perugia University. In 2007 he discussed the PhD thesis on Oriental Archaeology in Rome "La Sapienza" University. His main research activities are archaeological excavations and surveys in the Near East and studies on ancient Near Eastern religions, with an extensive interest in archaeoastronomy. He has been member of the Italian Archaeological Mission at Tell Mardikh-Ebla (Syria), under the direction of Prof. P. Matthiae, since 1998. From 2004 to 2007 he was also a member of the expedition to Khirbet al Batrawy (Jordan) and of the archaeological surveys on the Bronze Age monuments at Wadi az-Zarqa, directed by the Prof. L. Nigro.

**Table 1.** Orientation of 50 Nabataean monuments of a sacred character (temples, shrines, royal tombs and highplaces) as mainly obtained in December 2011. The table shows for each monument the location, the identification of the structure, the latitude and longitude (φ and λ), its azimuth (a) from inside looking out, and the angular height of the horizon (h) in that direction, and the corresponding declination (δ). The last column contains some additional comments or data for alternative orientations (in º).

| Place | Monument | φ (°/') | λ (°/') | a (°) | h (°) | δ (°) | Comments |
|---|---|---|---|---|---|---|---|
| Umm al Jimal | Nabataean temple | 32/20 | 36/22 | 18¼ | 0B | 52¾ | |
| Dhiban | Nabataean temple | 31/30 | 35/46 | 338 | 0½ | 52¼ | |
| El Qsar | Nabataean temple | 31/19 | 35/45 | 77¼ | 0b | 10½ | |
| Rabba | Roman temple | 31/16 | 35/44 | 109 | 0b | 16½ | Nabataean foundations? |
| Khir. Dharih | Nabataean temple | 30/54 | 35/42 | 194 | 6 | −50¼ | h~2½ N / δ~58 <br> h~13½ E / δ~−4¾ |
| Khir. Tannur | Nabataean temple | 30/58 | 35/42 | 92½ | 2 | −1¼ | |
| Dhat Ras | Nabataean temple I | 31/00 | 35/46 | 359½ | −0½ | 57¼ | h~−0½ E / δ~−0¼ |
| | Nabataean temple II | | | 267½ | −0½ | −2¾ | |
| | Nabataean temple III | | | 182 | 0½ | −58½ | Well preserved |
| Qsar Muhay | Central structure | 31/00 | 35/52 | 135½ | −0½ | −38¼ | Nabataean podium? |
| Wadi Ramm | Allat temple | 29/35 | 35/25 | 112 | 6½ | −15½ | |
| | East temple ? | | | 171¾ | 5½ | −54¼ | |
| Humayma | Nabataean temple | 29/57 | 35/21 | 90¾ | 1 | −0¼ | |
| Petra | Al Khazne (Treasury) | 30/20 | 35/27 | 65 | 17½ | 30 | Towards Siq <br> If h~13½° / δ~28½ |
| | Urn tomb | | | 264½ <br> 303 <br> 241½ <br> 295½ | 8 <br> 3 <br> 1½ <br> 4½ | −0¾ <br> 29½ <br> −23½ <br> 24 | ^ ^ at Umm al Biyara <br> To Ad Deir <br> SW-NE diagonal <br> NW-SE diagonal <br> 293^ ^296, see fig. 5 |
| | Corinthian tomb | | | 277 | 3½ | 7¾ | |
| | Palace tomb | | | 298 | 3 | 25½ | |
| | Sextus Florentinus | | | 344 | 6 | 61¼ | Roman period |
| | Khubtha H1 | | | 61 | 5 | 27¼ | H = Highplace |
| | Khubtha H2 | | | 259 | 3 | −8 | |
| | Khubtha H3 | | | 255 | 2½ | −11¾ | |
| | Khubtha H4 | | | 280 | 1½ | 9¼ | |
| | Lions' triclinium | | | 132¼ | 13 <br> 15 | −27 <br> −25½ | |
| | Ad Deir (Monastery) | | | 234 <br> 237½ | 8½ <br> 7 | −25¼ <br> −23½ | Exterior <br> Interior |
| | Structure 468 | | | 73¼ | 7½ | 18¼ | Deir Urn <br> a~80½ / h~8 / δ~12¼ |
| | Upper highplace | | | 296 | 2½ | 23½ | Near Deir <br> h~2½E / δ~−21 |
| | Lower highplace | | | 7½ | 1 | 59½ | Near Deir |
| | Deir altar | | | 157 | 5½ | −48 | |
| | Zibb Attuf | | | 92½ | 4¾ | 0¼ | |
| | Madbah (court) | | | 258 | 2½ | −9¼ | |
| | Madbah (altar) | | | 263 | 2½ | −6 | Djebel Harum <br> a~262 / h~3 / δ~−5½ |
| | Crescents' shrine | | | 252½ | 2 | −14¼ | |
| | Lion Mon. betyl | | | 154½ | 10 | −43 | |
| | Dushara shrine | | | 354½ | 7½ | 66½ | |
| | Garden temple | | | 156½ | 23½ | −31½ | |
| | Zib Faraum temple | | | 92¾ | 17½ | 6½ | |
| | Habis highplace | | | 270 | 7½ | 3¾ | |
| | Qsar al Bint | | | 18½ | 7½ | 61¼ | |
| | Qsar el Bint altar | | | 18½ | 9 | 62½ | |
| | QB Temenos | | | 289 | 16 | 24 | h~10½E / δ~−10¾ |
| | Winged Lion temple | | | 197 | 3 | −53¼ | |
| | Great temple (Civil use?) | | | 6½ | 5½ <br> 7 | 64¼ <br> 65¾ | Upper section <br> Lower section |
| | Peristyle temple | | | 19 | 7½ | 61 | |
| | Market temple | | | 6 | 7½ | 66¼ | |
| | Cardus | | | 97½ | 10 | −1¼ | |
| | East temple | | | 16½ | B | 55¼ | |
| Siq al Barid | "Khazna" | 30/22 | 35/27 | 147½ | 26½ | −25¼ | "Little Petra" |
| | Temple | | | 338 | 32> | 71 | |
| | Highplace | | | 227 | 12 | −28¼ | |
| Avdat | Nabataean temple | 30/48 | 34/40 | 62½ | 0 | 23 | |

**FIGURES and CAPTIONS**

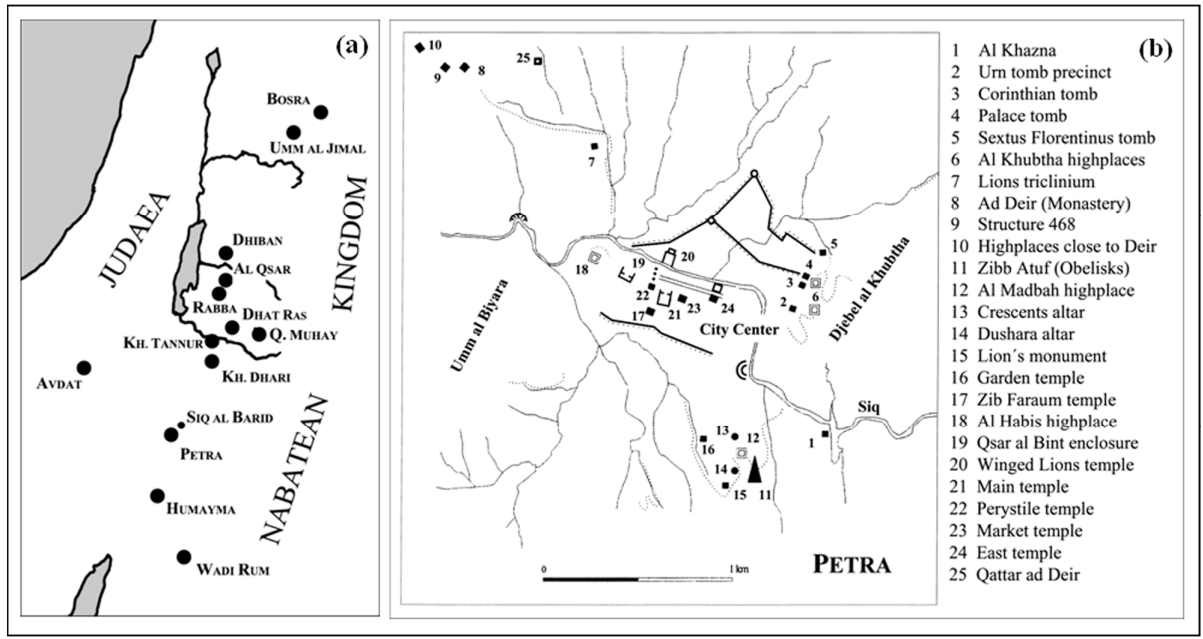

**Fig. 1.** Maps of the sites mentioned in the text and where the data have been collected. (a) Sites of the Nabataean Kingdom discussed in the text. (b) Sites of the city of Petra. Diagrams by the authors.

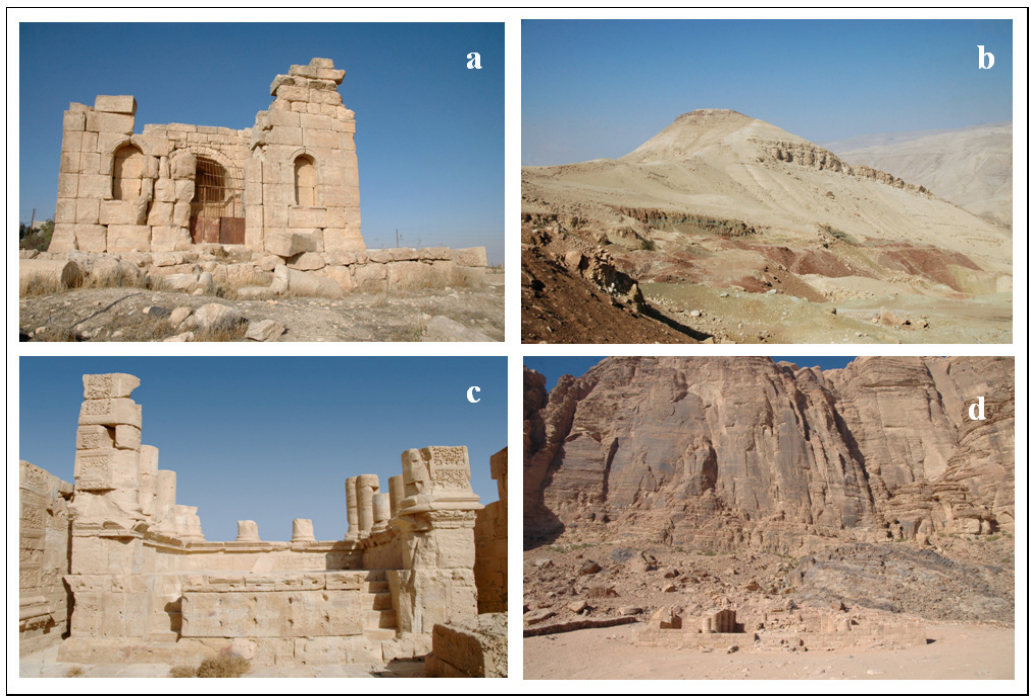

**Fig. 2.** Nabataean temples outside Petra: (a) Sanctuary at Dhat Ras; (b) Djebel Tannur, the temple is located at its summit; (c) Sancta sanctorum of the temple of Khirbet ed-Dharih; and (d) the temple of Allat dwarfed by the impressive cliffs of Djebel Ramm. Photographs by Juan A. Belmonte.

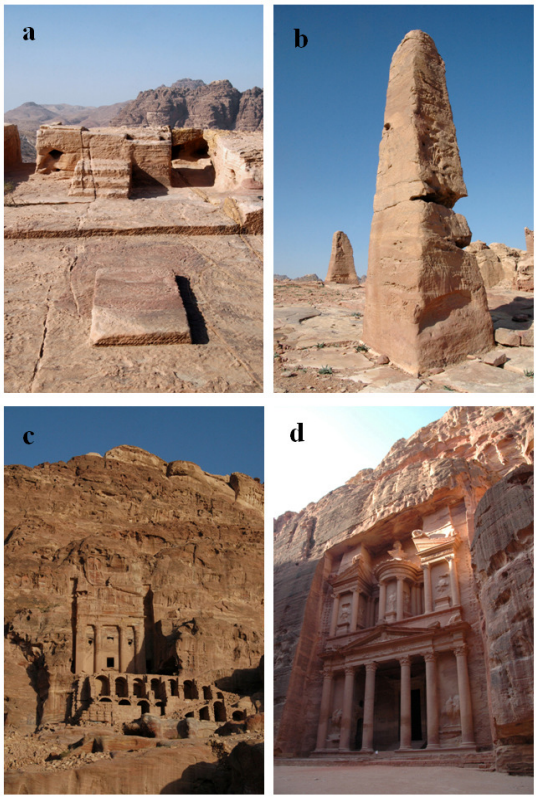

**Fig. 3.** Images of different kind of monuments of Petra measured, analyzed and discussed within the text: (a) Madbah highplace; (b) the Obelisks or Zibb Attuf at Djebel Madbah; (c) The urn Tomb at the cliffs of Djebel Khubtha; and (d) the splendid façade of Al Khazna. Photographs by J.A. Belmonte.

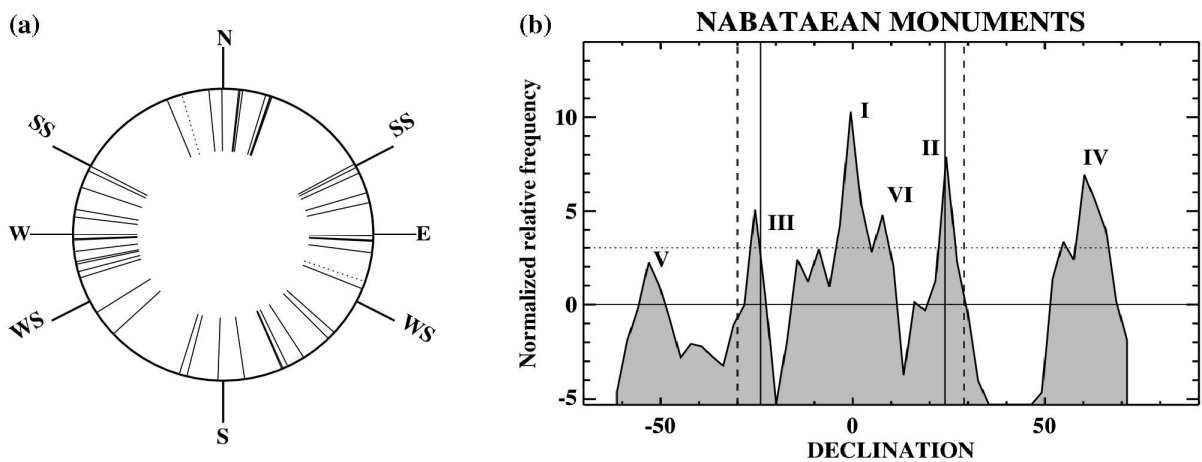

**Fig. 4.** Main outcomes of the archaeoastronomical analysis of Nabataean monuments: (a) orientation diagram of the structures (dot-lines for Roman period ones); (b) declination histogram of the set of monuments. Dashed and continuous vertical lines stand for major lunastices and solstices, respectively. Horizontal dot-line stands for the 3σ confidence level. Roman numbers identify peaks discussed within the text.

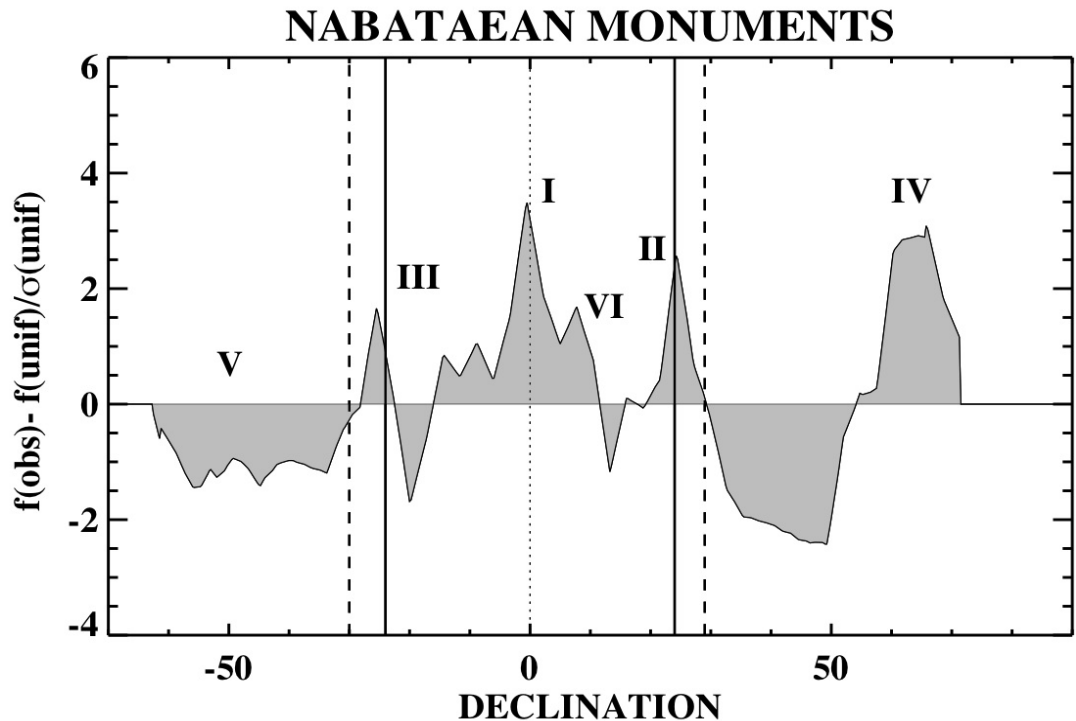

**Fig. 5.** Results yielded by a new analysis of the data in an attempt to test the significance of the peaks of accumulation. The observed declination histogram −f(obs)− is compared with the one that would be yielded by a uniform distribution of azimuths with the same number of data −f(unif)− and expressed in unit of the standard deviation of this second distribution. Peak IV is still significant within the distribution but peak V, which had already a low significance, looses all its weight.

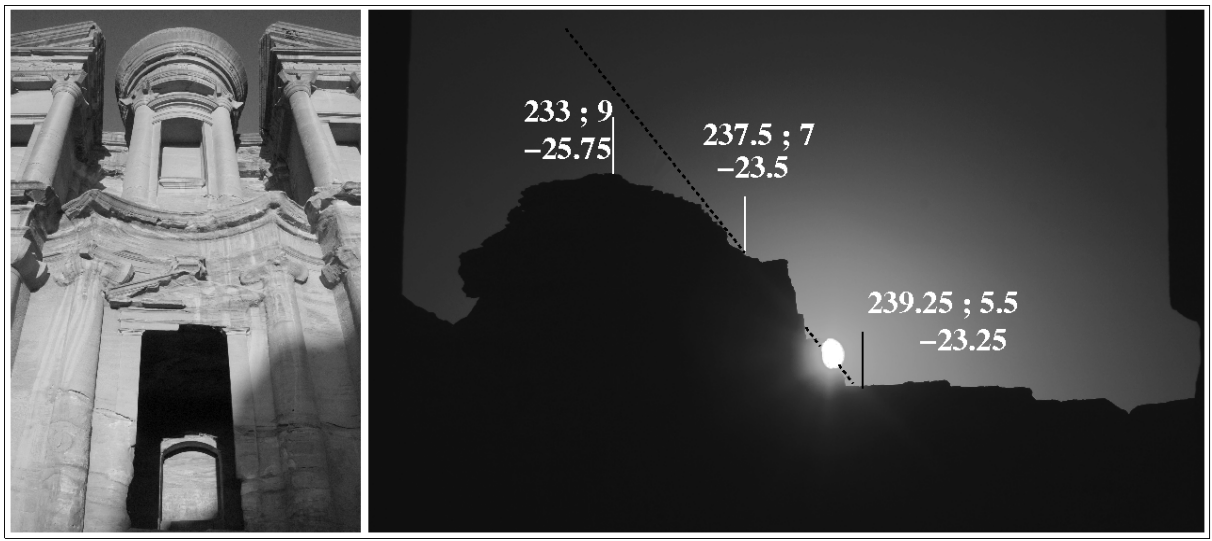

**Fig. 6.** Winter solstice sunset at Ad Deir. The left image shows the light and shadow effect in the innermost sacred area of the structure, the môtab. The right image demonstrates the accurate solstitial phenomenology associated with the site. Dotted line corresponds to the path followed during winter solstice sunset by the upper limb of the sun for the 1st century BC. Photographs by J.A. Belmonte and A.C. González-García, respectively.

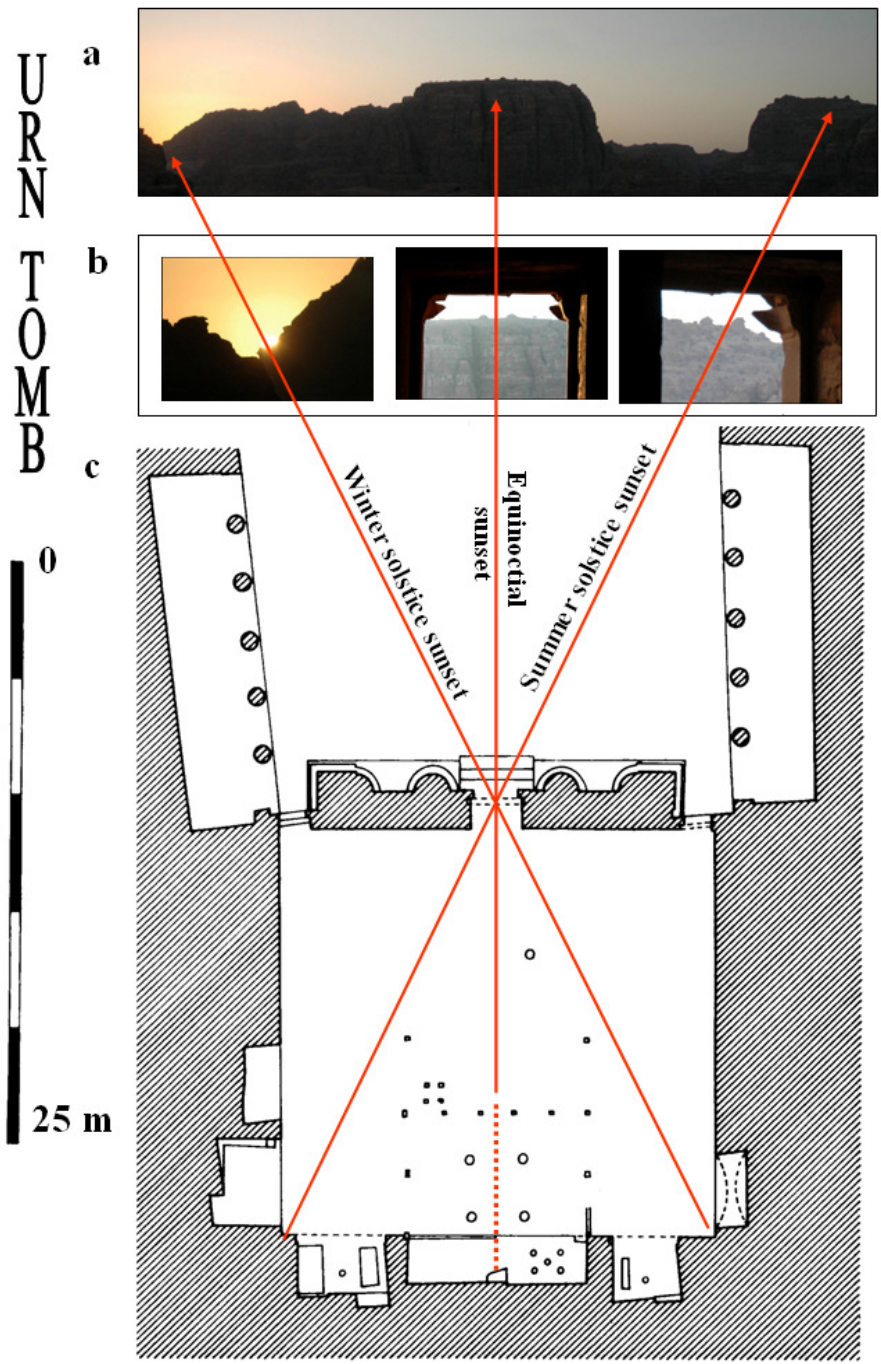

**Fig. 7.** Sunset phenomenology in the western horizon (a) related to the solstices and the equinoxes as seen from the Urn tomb enclosure (b). Our data suggests that the site and the internal distribution of the monument were deliberately chosen with an astronomical objective in mind. Photographs by J.A. Belmonte; Urn tomb plan (c) adapted from Guzzo and Schneider [1997].

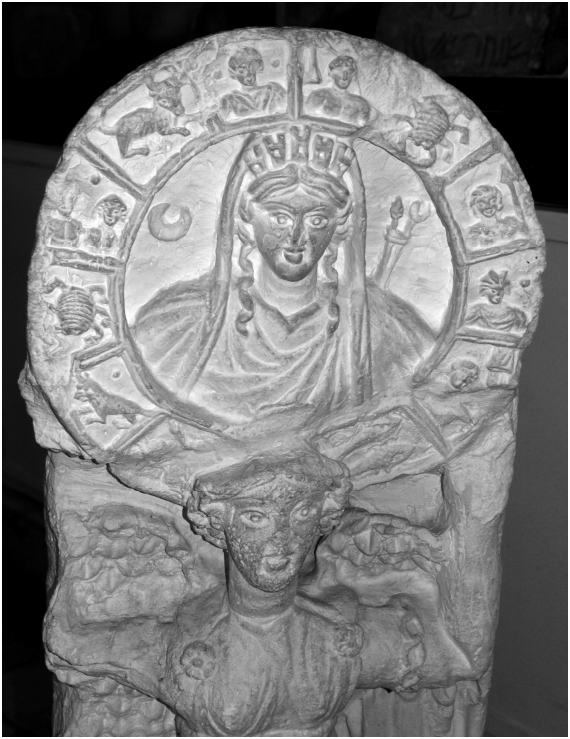

**Fig. 8.** A reconstructed plaster copy of the portrait of Tyche discovered at the temple of Tannur surrounded by the zodiacal signs. This is just one of the examples of astral symbolism in Nabataean remains. Original fragments at the Museums of Cincinnati and Amman. Photograph by Juan A. Belmonte, courtesy of the Amman Archaeological Museum.